\documentclass{iaus}
\usepackage{graphicx}

\def\arcsec{\hbox{$^{\prime\prime}$}}

\title[Planetary Nebula H$\alpha$ Fluxes] 
{A catalogue of integrated H$\alpha$ fluxes for $\sim$1100 Galactic planetary nebulae}

\author[I.\,S.~Boji\v{c}i\'c, D.\,J.~Frew \& Q.\,A.~Parker]   
{I.\,S.~Boji\v{c}i\'c$^{1,2,3}$, D.\,J.~Frew$^{1,2}$ and Q.\,A.~Parker$^{1,2,3}$}

\affiliation{$^1$Department of Physics and Astronomy, Macquarie University, Sydney, NSW 2109, Australia \\[\affilskip]
$^2$Macquarie University Research Centre in Astronomy, Astrophysics \& Astrophotonics\\[\affilskip]
$^3$Australian Astronomical Observatory, PO Box 296, Epping, NSW
1710, Australia\\email: {\tt ivan.bojicic@mq.edu.au}}

\pubyear{2011}
\volume{283}  
\pagerange{119--126}
 \jname{Planetary Nebulae: an Eye to the Future} \editors{Arturo Manchado, Letizia Stanghellini \& Detlef Sch\"onberner, eds.}
\begin{document}

\maketitle

\begin{abstract}
We present new determinations of the integrated H$\alpha$ flux for $\sim$1100 Galactic planetary nebulae measured from the Southern H-Alpha Sky Survey Atlas (SHASSA) and its northern counterpart, the Virginia Tech Spectral-Line Survey (VTSS).  This catalogue is the largest homogeneous database of its kind, tripling the number of currently available measurements. 
\end{abstract}
\firstsection 
\section{Introduction}

For a planetary nebula (PN), the integrated flux in a hydrogen Balmer line is analogous to the apparent magnitude of a star, and one of the most fundamental parameters that needs to be determined. However, most of the largest, evolved PNe in the sky are of low surface brightness and have poorly determined H$\beta$ or H$\alpha$ fluxes, if known at all.  Those that are published are heterogeneous and inconsistent (cf. Kaler 1983; Xilouris et al. 1996; Madsen et al. 2006). The situation is better for most of the brighter and more compact PNe in the Galaxy, with a large number of H$\beta$ fluxes compiled by Acker et al. (1992), based on the efforts of a large number of workers. Here we present new determinations of the integrated H$\alpha$ flux for $\sim$1100 PNe derived from the Southern H-Alpha Sky Survey Atlas (SHASSA; Gaustad et al. 2001) and the Virginia Tech Spectral-Line Survey (VTSS; Dennison, Simonetti \& Topasna 1998).  SHASSA provides narrowband (H$\alpha$ + [N II]) and continuum images of the entire southern sky.  The survey has a rather low spatial resolution (48\arcsec\ pixels) but it is continuum subtracted and accurately flux calibrated.  VTSS is a northern H$\alpha$ survey, but remains uncompleted.  The resolution is 96\arcsec, which translates to a brighter flux limit due to confusion noise.

\begin{figure}
\begin{center}
\includegraphics[scale=0.36]{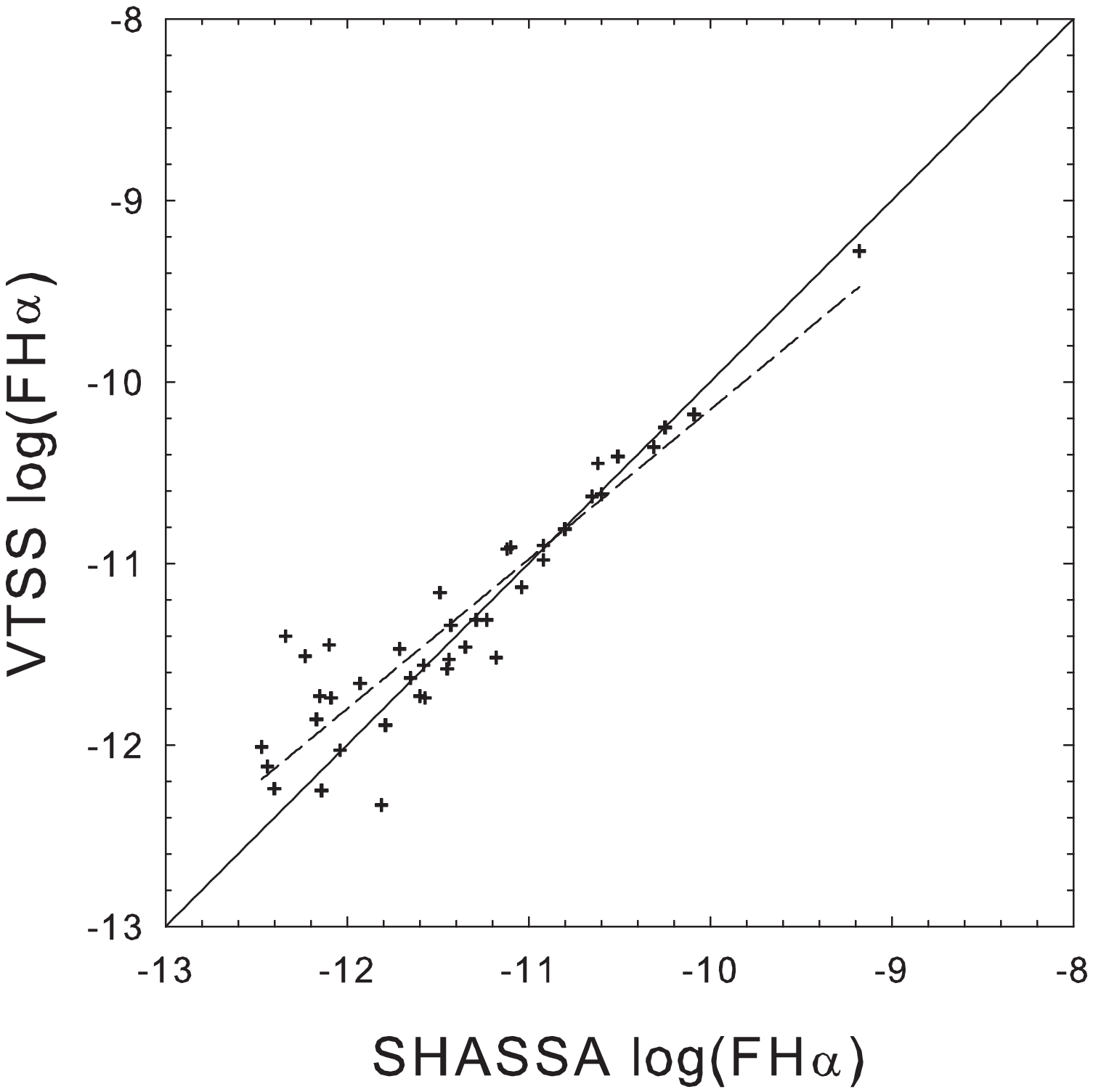}
\includegraphics[scale=0.36]{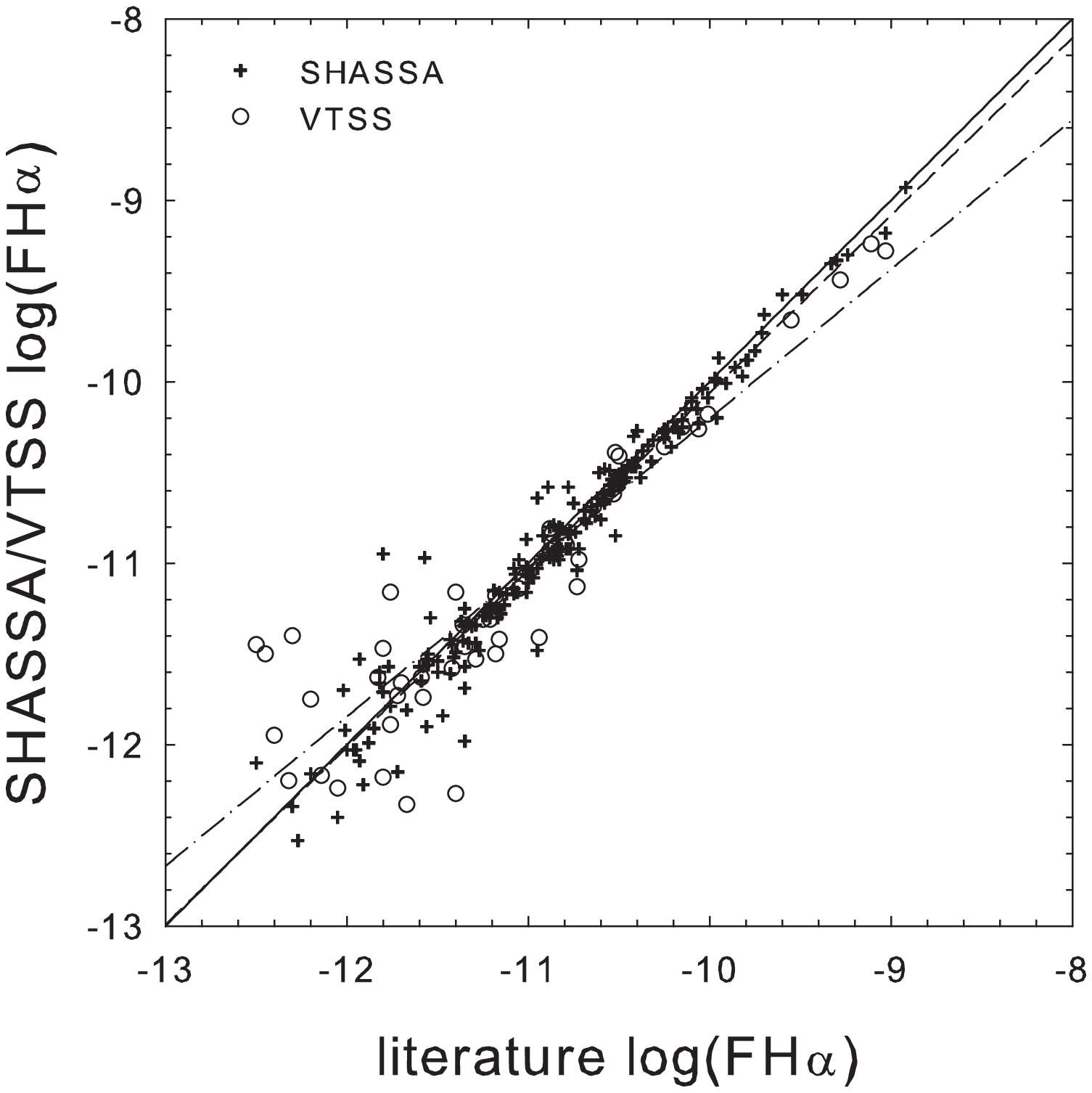}
\caption{A comparison of fluxes of PNe detected in  the SHASSA/VTSS overlap region (left) and combined SHASSA/VTSS against fluxes from the literature. Solid line: the expected 1-1 relationship; dashed and dash-dotted line: best fit to the SHASSA and VTSS data, respectively.}\label{figs}
\end{center}
\end{figure}

\section{Methodology and Results}
To make the measurements of the H$\alpha$ integrated fluxes for all the PNe reported here, the IRAF $phot$ task was used. The photometry pipeline was scripted in PyRAF, a Python interface to IRAF. For each object, a circular aperture was placed over the catalogued PN position. The diameter of the aperture was calculated from the point spread function value and the angular diameter of the nebula. Measurements affected by poor sky estimates, confusion with bright sources, or various artefacts are excluded from the catalogue and are re-measured manually.  Finally, the SHASSA red flux is corrected for the response of the H$\alpha$ filter which also passes both [NII] lines ($\lambda6548$ and $\lambda6584$), using values for the [NII]/H$\alpha$ ratio taken from the literature or our own spectroscopic data (Frew 2008).  

After weeding out probable mimics, following the recipe of Frew \& Parker (2010), we counted $\sim$2500 {\it bona-fide} PNe catalogued by Acker et al. (1992), Kohoutek (2001) and the MASH project (Parker et al. 2006, Miszalski et al. 2008) in the areas covered by SHASSA and VTSS.  From this total, we counted $\sim$1000 PNe detected in SHASSA and 160 PNe detected in VTSS ($\sim$1100 objects in total because of the overlap between two surveys).  PNe were assumed to be detected if: (i) at least one pixel within the aperture has a measured flux above sky + 5$\sigma$, and (ii) if rule (i) applies in more than 50\% of fields containing the PN. We examined the spatial distribution of both detected and non-detected PNe, and compared the measurements for PNe covered in both surveys, and between our new measurements and data available in the literature (Fig.~1).

The new SHASSA/VTSS catalogue of H$\alpha$ photometric fluxes for Galactic PNe (Frew et al. 2011, in prep.) will be the largest homogeneous database of its kind, tripling the number of currently available measurements. Calculations involving the integrated PN surface brightness, the ionized nebular mass, the rms electron density, the Zanstra temperature of the central star (and hence its luminosity and mass), and the PN luminosity function are all dependent on accurate integrated line fluxes, which this dataset will provide.  Also, in combination with our catalogue of radio flux measurements (Boji\v{c}i\'c et al. 2011a, Boji\v{c}i\'c et al. 2011b, in preparation), we will be able to independently derive robust interstellar extinction values for a large sample of PNe.  

\section*{References}
{\small
\noindent 
Acker, A., et al. 1992, SECGPN, ESO, Garching\\
Boji\v{c}i\'c, I. S., Parker, Q.A., Filipovi\'c, M.D. \& Frew, D.J. 2011, MNRAS, 412, 223\\
Dennison, B., Simonetti, J.H. \& Topasna, G.A. 1998, PASA, 15, 147\\
Frew, D.J. 2008, PhD Thesis, Macquarie University\\
Frew, D.J. \& Parker, Q.A. 2010, PASA, 27, 129\\
Gaustad J. E., et al. 2001, PASP, 113, 1326\\
Kaler, J.B., 1983, ApJ, 264, 594\\
Kohoutek, L. 2001, A\&A, 378, 843\\
Madsen, G.J., Frew, D.J.,  Parker, Q.A.,  Reynolds, R.J. \& Haffner, L.M. 2006, IAUS, 234, 455\\
Miszalski, B., et al. 2008, MNRAS, 384, 525\\
Parker, Q.A., et al. 2006, MNRAS, 373, 79\\
Xilouris K. M., et al. 1996, A\&A, 310, 603
}

\end{document}